\documentclass[prl,10pt,twocolumn,amsmath,amssymb,showpacs]{revtex4}

\usepackage{graphicx}

\begin{document}
\title{Hexatic and mesoscopic phases in the 2D quantum Coulomb system}

\author{Bryan K. Clark,$^{1}$ Michele Casula,$^{2}$ and D. M. Ceperley$^{1,3}$}

\address{ $^1$ Department of Physics, University of Illinois at Urbana-Champaign,
1110 W. Green St, Urbana, IL 61801, USA\\
$^2$ Centre de Physique Th\'eorique, Ecole Polytechnique, CNRS, 91128 Palaiseau, France \\
$^3$  NCSA, University of Illinois at Urbana-Champaign, Urbana, IL 61801, USA }
\date{\today}
\begin{abstract}
We study the Wigner crystal melting in a two dimensional
quantum system of particles interacting via the $1/r$ Coulomb
potential. We use quantum Monte Carlo methods to calculate its
phase diagram, locate the Wigner crystal region, and analyze
its instabilities towards the liquid phase.  We discuss the
role of quantum effects in the critical behavior of the system,
and  compare our numerical results with the classical theory of
melting, and the microemulsion theory of frustrated Coulomb
systems. We find a Pomeranchuk effect much larger then in solid
helium. In addition, we find that the exponent for the
algebraic decay of the hexatic phase differs significantly from
the Kosterilitz-Thouless theory of melting.  We search for the
existence of mesoscopic phases and find evidence of metastable
bubbles but no mesoscopic phase that is stable in equilibrium.
\end{abstract}
\pacs{29.25.Bx. 41.75.-i, 41.75.Lx}

\maketitle

The Wigner crystal (WC) melting has been a subject of intense study over the years.\cite{strandburg,waintal,manousakis}
A better understanding of this process is particularly important in the two dimensional (2D)
one component plasma (OCP) with $1/r$ Coulomb interactions,
since it could explain many features
in systems such as electrons at interfaces, dusty plasmas, MOSFETs,
and charged colloids.

Although in 2D, a true crystalline order is not possible at any
finite temperature T \cite{jancovici}, a quasi-long range
translational order is stable at strong coupling. Upon melting
from the WC a variety of mechanisms have been
proposed.\cite{strandburg} At high temperature, Halperin and
Nelson \cite{nelson} suggested that the 2D melting is a two-step process,
with a hexatic phase in between.
Hexatic phases have been seen in colloids, liquid crystals, and
dusty plasmas,\cite{experiments} and found in some classical
simulations.\cite{classical_hexatic,classical2}
However, there is disagreement about its location, critical
exponents,  and order. The influence of quantum effects is
unclear.

In the ground state, Jamei \emph{et al.} \cite{Kivelson} argue by a mean-field
approach that a first order WC-liquid transition at zero
temperature is disallowed; instead the transition is mediated
by microemulsion phases (e.g. stripes or
bubbles).
This mechanism might be related to the stripe order in some
strongly correlated materials and anomalous effects in
experiments on MOSFETs.\cite{metal_insulator}

In this Letter, we examine a 2D system of N quantum
distinguishable (Boltzmannon) particles that interact with a
Hamiltonian:
\begin{equation}
\label{hamiltonian} H=-\frac{1}{r_s^2}\sum_i \nabla_i^2 +
\sum_{i<j} \frac{2}{r_s|\mathbf{r}_{i}-\mathbf{r}_{j}|} + V_\textrm{bg},
\end{equation}
where $\mathbf{r}_{i}$ is the location of the $i$'th particle.
The system is parameterized by two dimensionless parameters which
we choose to be $r_s$ and $T$.   
We use $a=r_s a_0$ as the unit of length, with $a_0$ the Bohr radius, Rydberg as
the unit of energy, and temperature; the density is $1/\pi
(r_s  a_0)^2$. The system is neutralized by a rigid homogeneous
background, which gives rise to the constant $V_\textrm{bg}$.
Our goal is to map out the phase diagram, 
to locate the WC phase boundaries, and analyze the
transition to the liquid. To do so, we use quantum Monte Carlo
(QMC) techniques, which are uniquely suited for calculations of
the properties of strongly coupled Coulomb systems.

At finite temperature, we use path integral Monte Carlo
(PIMC),\cite{PIMC} to sample configurations from the thermal
density matrix. 
In contrast to fermion systems, for a system of distinguishable
particles, there is no sign problem. All systematic errors,
such as the time step and finite size errors, can be studied
with extrapolation. \cite{timestep} Our calculations involve
approximately 100 phase points at system ranging from 200 to
2248 particles and between 10 and 8000 slices in imaginary
time.

At $T=0$, we use diffusion Monte Carlo (DMC), a projector
method that takes as input a trial wave function $\Psi_T$, and
filters from it, all excited states. Calculating DMC energies
in this system is formally exact as long as $\Psi_T$ has a
non-zero overlap with the ground state.

\begin{figure}[htp]
\includegraphics[width=\columnwidth]{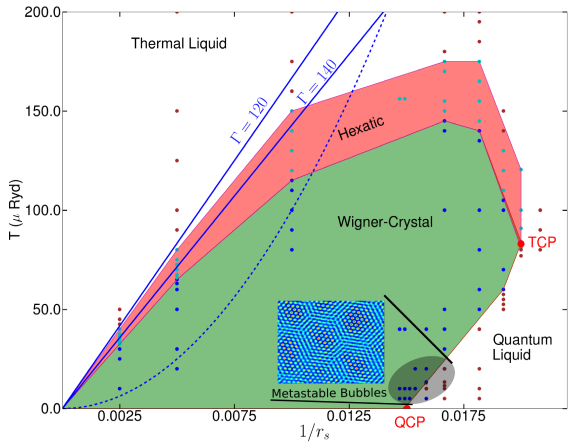}
\caption{Phase diagram of the quantum 2D Coulomb system
calculated using QMC as obtained by analyzing correlation
functions. Points indicate location of QMC calculations. The
phase boundaries are interpolations between these points. The
dotted blue line indicates a rough separation between the classical and the quantum
regions: when the thermal de Broglie wave length equals the
interparticle spacing. The solid straight blue lines are the
classical results of Ref.~\cite{classical2}. The inset is a
density profile of the metastable bubbles seen in PIMC.
}\label{fig:phasediagram}
\end{figure}

\textit{General properties of the phase diagram -} Our proposed
phase diagram is shown in Fig.~\ref{fig:phasediagram}. The
boundary between the WC and the liquid phase presents a clear
re-entrant behavior at low $T$. Although such behavior has been
seen in 
helium,\cite{Reentrant_Helium_1,Reentrant_Helium_2} for the
2D-OCP the effect is significantly stronger. We find the ratio
between the minimal and maximal density of the crystal $0.6$
compared to $0.03$ for He$^3$ and $3 \cdot 10^{-5}$ for He$^4$.
The re-entrant liquid terminates in a zero temperature quantum
critical point (QCP)
at $r_s \simeq 66.5\pm 0.2$.
This differs by 10 \% from previous DMC
simulations;\cite{BosonGroundState} our calculations did a
more careful extrapolation to the thermodynamic limit.\cite{chiesa}

At large $r_s$ we compare our results with simulations done for
the classical OCP;  
classical simulations found a WC phase stable for $\Gamma > 140$ and a liquid phase stable for $\Gamma < 120$
,\cite{classical2} where $\Gamma=2/r_s T$. Seen as lines
in Fig.~\ref{fig:phasediagram}, these values compare favorably
with our results at large $r_s$. For $120 \leq \Gamma \leq
140$, we found a hexatic quasi-long range order; this order
extends into the quantum regime by bending with respect to the
classical line as the density is increased (i.e. for $r_s <
200$).

\begin{figure}[htp]
\includegraphics[width=\columnwidth]{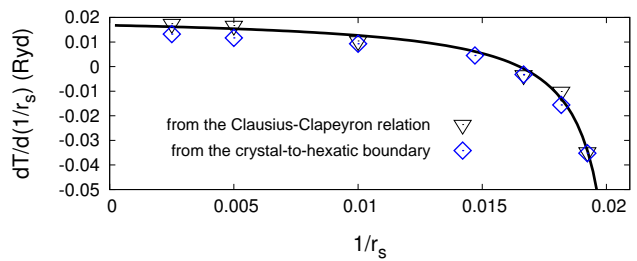}
\caption{$dT/d(1/r_s)$ vs $1/r_s$ computed using the
Clausius-Clapeyron relation, Eq.~\ref{claformula} for $N=562$,
compared with the results obtained directly from the phase
diagram in Fig.~\ref{fig:phasediagram} for the first order
WC-to-hexatic boundary.} \label{fig:clacla} 	
\end{figure}

At $r_s \simeq 60$ the melting temperature decreases until it
reaches  a tricritical point (TCP), located at $r_s \simeq 50$
and $T \simeq 80 ~ \mu Ryd$, where the isotropic liquid, the
hexatic phase, and the WC coexist. We have used the
Clausius-Clapeyron relation  to understand this behavior. For
Coulomb systems it reads\cite{david}:
\begin{equation}
\frac{dT}{d(1/r_s)}=\frac{2}{\Gamma} \frac{\delta K + \delta U}{\delta U},
\label{claformula}
\end{equation}
where $\delta U = U_c - U_h$ and $\delta K = K_c - K_h$ are the
changes in the internal and kinetic energies between the
crystal $(U_c,K_c)$ and hexatic $(U_h ,K_h)$ phase. The slope
is shown in Fig.~\ref{fig:clacla}. At the downturn, the slope
is zero, while it diverges at the TCP, where the entropy
difference, and hence $\delta U$, goes to zero. In fact,
Eq.~\ref{claformula} shows that a first order line for any
Coulomb system, must become second order at the nose of a
re-entrant phase.

\begin{figure}[htp]
\includegraphics[width=\columnwidth,height=50mm]{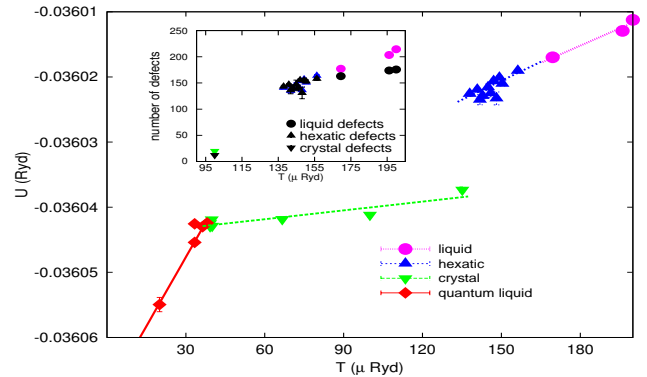}
\caption{
Internal energy per particle versus temperature at $r_s=55$ for
$N=562$.
The inset plots both the number of defects and the
internal energy scaled to arbitrary units and shifted.} \label{fig:Energy}
\end{figure}

There are two WC-to-liquid boundaries: the quantum melting
line, going from the QCP to the TCP, and the thermal melting
line, flowing from the high temperature side of the TCP towards
lower density (large $r_s$), and characterized by a two-step
process mediated by a hexatic phase. The order of the phase
transitions is established by examining the internal energy 
per particle
$U=U(T)$ at a series of densities $r_s \in (48,400)$ through a
wide range of temperatures. The behavior at $r_s=55$ is shown
in Fig.~\ref{fig:Energy}. A second order (or weakly first
order) phase transition is evident for the quantum melting
line, indicated by a kink in the internal energy when moving
from the re-entrant liquid phase to the WC. Increasing the
temperature, $U(T)$ shows a sharp jump at the boundary between
the solid and hexatic phase, indicating a first order
WC-hexatic transition. 
The internal energy is continuous between the hexatic phase and the liquid, indicating a continuous hexatic-liquid transition.
This is in disagreement with
the classical theory of melting, which suggests two continuous
transitions of Kosterlitz-Thouless (KT) type. It should
be noted that as one approaches the classical limit, $r_s$
($\approx 400$), the jump in energy shrinks and becomes barely
distinguishable from a continuous transition.

\begin{figure}[htp]
\includegraphics[width=\columnwidth]{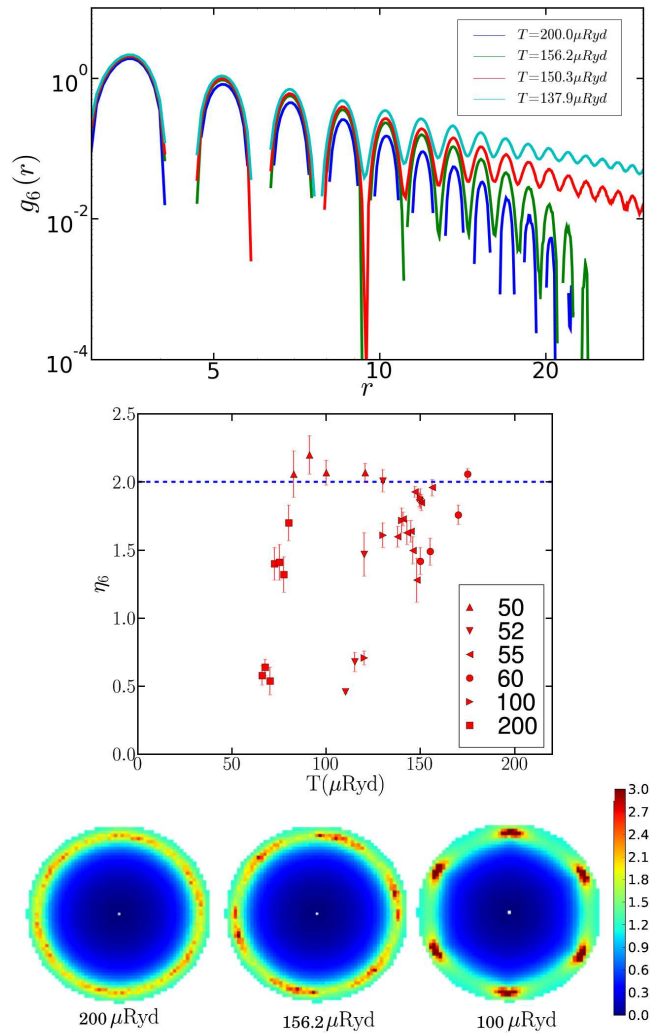}
\caption{ Top: $g_6(r)$ at different temperatures for $r_s=55,
N=2248$.  The lower two temperatures 
show algebraic decay while the higher temperatures decay
faster than algebraically. 
Middle: $\eta_6(T)$ in the hexatic phase for $N=562$ at various
temperatures and densities (for $r_s$ values, see legend on figure).
All exponents satisfy $\eta_6 \lesssim 2$ (shown as dotted line).
Bottom: $S(\mathbf{k_x},\mathbf{k_y})$
in a liquid, hexatic, and solid respectively for $r_s=55, N=562$, 
temperatures in caption.}
\label{fig:hexatic}
\end{figure}

\textit{Thermal melting line -} The thermal melting line is
characterized by the presence of a hexatic intermediate phase
with an order parameter:
\begin{equation}
g_6(r)=\left< \sum_{ij} \Psi^*(\mathbf{r}_i)\Psi(\mathbf{r}_j)\delta((|\mathbf{r}_i-\mathbf{r}_j|)-r) \right>,
\end{equation}
where $\Psi(\mathbf{r}_i)=\frac{1}{6}\sum_{\langle j \rangle}
\exp(6i\theta_{ij})$; the summation is over nearest neighbor
particles $j$ surrounding $i$, \cite{nn} and
$\theta_{ij}$ is the angle between an arbitrary fixed vector
and the vector joining $\mathbf{r}_i$ to $\mathbf{r}_j$. The algebraic
decay of $g_6(r)$ signals a quasi-long range orientational
order and this, combined with a lack of translational order,
defines the hexatic phase. Additionally, the orientational
order of the hexatic phase manifests itself via a sixfold
anisotropy in the structure factor $S(\mathbf{k})$. Fig.~\ref{fig:hexatic} shows
the behavior of the $S(\mathbf{k})$ in the liquid, solid, and hexatic
phases at $r_s=55$, and  $g_6(r)$ as a function of temperature.
There is a clear qualitative change
in the decay of $g_6(r)$ for $156.2~\mu Ryd \leq T \leq 150.3~\mu Ryd$,
as we transition from the liquid to the hexatic phase.

The KT theory predicts that for hexatic systems in the XY
universality class $g_6(r) \propto r^{-\eta_6}$, with $\eta_6
\le 1/4$, 
and $\eta_6$ reaching its maximum $1/4$ at the
transition to the liquid.
We fit the large distance values of $g_6(r)$, to determine
${\eta_6}$ and found that, independent of temperature, it
undergoes the hexatic-to-liquid transition for $\eta_6 \simeq 2$.
The disagreement with the KT theory is likely due to the $1/r$
interaction. Indeed, critical exponents in other models, for
example the Ising model, are known to be strongly affected by
the range of the interaction.\cite{urbana_paper}

Defects play an important role in 2D melting. We define defects
using a Voronoi construction of each particle's centroid.
In a perfect hexagonal lattice, each centroid has six
neighbors. 
Defective centroids, say with 5 or 7 neighbors, are denoted as
vortices and antivortices respectively. According to the KT
theory the WC-hexatic transition is caused by the unbinding of
dislocations (particles defined as strongly bound
vortex-antivortex pair), while the hexatic-liquid transition is
driven by the vortex-antivortex unbinding. However, important
differences are found in our system. At the verge of the WC
melting, instead of the expected continuous rise of
dislocations, the number of defects  jumps discontinuously. We
see a strong correlation between the energy and the number of
defects;
$U(T)$ in the hexatic and solid phases is controlled entirely
by their number. Therefore, the first order jump seen in $U(T)$
at the WC-hexatic transition is due to the proliferation of
dislocations in the system, which can be induced by grain
boundaries,\cite{fisher,chui} present in the WC phase. The
inset of Fig.~\ref{fig:Energy} shows this proportionality by
plotting both the defect number and the energy of the various
phases on the same axis, with the  $U(T)$ scaled to arbitrary
units and shifted. We note that the first order transition in a
long range system is unconventional because the presence of the 
background does not allow macroscopic separation of phases with
different densities.
In the hexatic-liquid transition, the number of defects grows
slowly while the energy of the system rises at a much faster
rate suggesting the presence of an unbinding process possibly
related to the vortex-antivortex melting.\cite{nelson}

\textit{Quantum melting line -} The quantum melting line goes
from the QCP to the TCP.
Jamei \emph{et al.}\cite{Kivelson} and Ortix \emph{et
al.}\cite{dicastro} showed recently by a mean-field Hamiltonian
that the system undergoes a WC-liquid transition at $T=0$
through a sequence of intermediate phases such as stripes,
or other ``microemulsion phases''.
Their shape and size depend on the parameters of the mean-field Hamiltonian. 
At the critical density $r_s^c$, where the free energy of the
crystal equals that of the liquid, the optimal geometry  is
found to be alternating liquid and crystal stripes of width
$W_0= a \exp(\lambda)$, with $a$ a lattice cutoff proportional
to the mean interparticle spacing, $\lambda = 4 \pi^2 e^2
\sigma / \Delta\mu_c^2$,
where $\sigma$ the surface tension, and $\Delta\mu_c=
\mu_\textrm{crys}(r_s^c)-\mu_\textrm{liq}(r_s^c)$ the chemical
potential difference between the crystal and liquid at
criticality.

At finite $T$, our PIMC results suggest a second order (or very
weakly first order) direct transition. We carefully searched
for mesoscopic phases in the surrounding region.  
At $r_s \approx 55$, where the transition temperature to the
re-entrant liquid is relatively high ($T \simeq 50 ~ \mu Ryd$),
we find no evidence for mesoscopic phases. At larger $r_s$ and,
hence, lower transition temperature, the PIMC simulations
occasionally yield inhomogenous phases. A density profile of
such a phase is shown in the inset of
Fig.~\ref{fig:phasediagram}. However, those inhomogenous
structures appear to be metastable because their internal
energy, measured in PIMC at low T ($< 10~ \mu Ryd$) and DMC at
T=0, is always higher than the homogenous phases. To clarify
this situation, we performed additional DMC simulations to
determine the parameters of the mean-field model. In the liquid
phase, we used a pair product or Jastrow trial function
$\Psi_{\text{liquid}}=\prod_{i<j} \exp [- u_{RPA}(|\textbf{r}_i - \textbf{r}_j|)]$, 
with $ u_{RPA}$ the RPA Jastrow function.\cite{david_rpa}  
For the
solid and stripe phase we multiplied by a Gaussian, tieing the
distinguishable particles to predefined sites, $\{\textbf{I}_i\}$
arranged to correspond to a crystal or stripe: 
$\Psi_{\text{crystal}}=\Psi_{\text{liquid}} \exp[-\sum_{i}
\alpha |\textbf{r}_i-\textbf{I}_i|^2]$.
We fit the $T=0$ liquid and crystal energies to a polynomial
and, assuming the transition is weakly first order, obtain
$r_s^c=66.5\pm 0.2$ and $\Delta\mu_c = 65\pm5 ~\mu Ryd$.

We estimate the surface tension $\sigma$ by computing
the energy at $r_s^c$ for different number of
stripes,\cite{overlap} and
hence different surface length in the system:
$\sigma_\textrm{stripe}=1.55\pm 0.09 ~ \mu Ryd / a_0$. Plugging
$\sigma_\textrm{stripe}$ and $\Delta\mu_c$ in the equation for
$\lambda$
gives $\lambda \approx 3 \times10^5$, which leads to the width of the
stripes at the critical point of $W_0 \approx 10^{10^5}$: much
larger than any physical system. The large widths are a result
of very small difference in the chemical potential between the
homogeneous liquid and crystal phases at the QCP,
characteristic of the $1/r$ interaction.
Away from $r_s^c$, the periodicity of the alternating stripes will become even larger.
Therefore, the
instability towards
``microemulsion phases'' has a characteristic emergent length
$W_0$ so large that it cannot explain non-Fermi liquid behavior
seen in 2D experimental setups.\cite{metal_insulator}

\textit{Conclusion - }  We have studied the WC melting in the
2D quantum Coulomb system of particles with Boltzmann
statistics. Its phase diagram shows thermal and quantum
melting. The thermal melting is mediated by a hexatic phase,
but significant deviations from the classical KT theory are
found. The WC-hexatic transition is first order, driven by the
proliferations of defects in the crystal, where they are
assembled into grain boundaries. The hexatic parameter $\eta_6
\approx 2$ belongs to the universality class of long-range
models and reveals the importance of the long-range interaction
in determining the transition properties. In the low
temperature region, we found
a strong Pomeranchuck effect, much larger than for solid
helium. At low temperatures, we did not find stable
microemulsion phases. An estimate of their size is exceedingly
large, 
which makes this kind of phases impossible to see in any
physical system if driven by correlation effects only.

We acknowledge support from NSF grant DMR-0404853 and
EAR 05-30643. MC also
thanks the Centre de Physique Th\'eorique of the Ecole
Polytechnique.
Supercomputer calculations were done at the NCSA. We would also like to thank also Frank Kruger for his careful reading and John A. Goree and Yu. E. Lozovik for useful discussions.


\begin{thebibliography}{00}
\bibitem{timestep} Most of the path integral
calculations were done using the pair action approximation of
the diagonal and with a time step $\tau=50$. A study of time
step error indicates that the forms of the phases as well as
differences in energy have converged.
\bibitem{nn} Nearest neighbors to
$i$ are all particles $j$ with $|\mathbf{r}_i-\mathbf{r}_j| <2.54$.
\bibitem{overlap} We exploit the small overlap between
the liquid and the crystal wave function to project out the lowest
energy in the Hilbert space within the same ``phase'' of the trial function.
\end{thebibliography}
\end{document}